\documentclass[conference]{IEEEtran}
\IEEEoverridecommandlockouts
\usepackage{times}
\usepackage{graphicx}
\usepackage{authblk,cite,xspace,url,setspace}
\usepackage{float, rotating}
\usepackage[cmex10]{amsmath}
\usepackage{amsthm}
\usepackage{amsmath}
\usepackage{algorithmic}
\usepackage{array}
\usepackage{bm}

\usepackage[tight,footnotesize]{subfigure}
\usepackage{url,caption, color, enumerate}

\title{When Wireless Security Meets Machine Learning: Motivation, Challenges, and Research Directions}

	\author[1]{Yalin E. Sagduyu}
	\author[1]{Yi Shi}
	\author[1]{Tugba Erpek}
	\author[2]{William Headley}
	\author[3]{Bryse Flowers}
	\author[4]{George Stantchev} 
	\author[5]{Zhuo Lu}
	\affil[1]{\normalsize Intelligent Automation, Inc., Rockville, MD 20855, USA}
	\affil[2]{\normalsize Virginia Tech, Blacksburg, VA 24060, USA}
	\affil[3]{\normalsize University of California, San Diego, CA 92093, USA}
	\affil[4]{\normalsize Naval Research Laboratory, Washington, DC 20375, USA}
	\affil[5]{\normalsize University of South Florida, Tampa, FL 33620, USA}
	
\begin{document}
\newcommand{\argmax}{\arg\!\max}
\maketitle

\begin{abstract}
Wireless systems are vulnerable to various attacks such as jamming and eavesdropping due to the shared and broadcast nature of wireless medium. To support both attack and defense strategies, machine learning (ML) provides automated means to learn from and adapt to wireless communication characteristics that are hard to capture by hand-crafted features and models. This article discusses motivation, background, and scope of research efforts that bridge ML and wireless security. Motivated by research directions surveyed in the context of ML for wireless security, ML-based attack and defense solutions and emerging adversarial ML techniques in the wireless domain are identified along with a roadmap to foster research efforts in bridging ML and wireless security.    
\end{abstract}

\section{Introduction}
\subsection{Motivation}
The research area of radio frequency machine learning (RFML) has had extremely strong growth in recent years. RFML solutions have been proposed to solve many problems in the areas of wireless communications, networking, and signal processing such as cognitive radio, spectrum sensing, spectrum coexistence, jamming/anti-jamming, emitter identification, and intrusion detection \cite{WirelessDL}. While machine learning (ML) is getting traction in wireless security applications such as detecting conventional attacks, little work has been done so far in investigating how vulnerable wireless systems are to ML-based security and privacy attack vectors that have recently been considered in other modalities such as image recognition and natural language processing. In addition, commercial applications for the internet of things (IoT) and 5G communications and recent interest by government entities in defense mechanisms to protect ML applications have shown that security and privacy concerns of ML systems are extremely timely and relevant.

\subsection{Scope and Background}
	ML offers invaluable tools for a diverse and far-reaching set of applications ranging from image recognition and natural language processing to cyber security and autonomous navigation. In recent years, applications of ML have emerged in the wireless communications domain, forming a major ingredient of a more general topic area, colloquially referred to as RFML. In particular, ML systems based upon state-of-the-art deep learning architectures, powered by the ever-increasing hardware accelerations for computing, have been considered for spectrum sensing applications (signal detection, estimation, classification, and identification), channel estimation, emitter identification, cognitive jamming and anti-jamming, among many others. The success of these research thrusts are expected to play an increasingly prevalent role in the development of future generations of commercial wireless networks \cite{Letaief2019}.  
	
	In the more established ML domains, recent research has demonstrated the efficacy of utilizing adversarial ML to negatively impact the performance of ML systems. Additionally, vulnerabilities to the privacy and security of these systems, and the data used to train the systems, have been exposed. However, the impact of these concepts to RFML technologies is currently underdeveloped. Therefore, it is a timely research effort to investigate the interaction of RFML with wireless security, privacy, robustness, and resilience. Given these facts, it is imperative to  combine research efforts in ML, RFML, privacy, security, and wireless communications to further advance the state-of-the-art in security techniques, architectures, and algorithms for ML in wireless systems.
	
	Research efforts that are needed to address ML for wireless security are diverse and include (but not limited to) the following directions:
	
	\begin{itemize}
	\item \emph{Adversarial ML Techniques}: evasion attacks (adversarial examples), poisoning (causative) attacks, Trojan (backdoor or trapdoor) attacks, generative adversarial learning.
	\item \emph{Privacy Issues of ML Solutions}: membership inference attacks, model inversion, physical layer privacy.
	\item \emph{ML Hardening Solutions}:  privacy-preserving learning, secure learning, hardware and software implementations, edge computing, testbeds and experiments, datasets.
	\item \emph{Relevant ML Applications for Wireless Security}: device identification, spectrum monitoring, RF fingerprinting, smart jamming, smart eavesdropping, localization, covert communication, authentication, anonymity, intrusion detection, IoT security.
	\end{itemize}

This article surveys the recent efforts to address ML for wireless security in Section \ref{se:research_efforts}, documents the challenges that are still left to solve in Section \ref{se:challenges_and_gaps}, and concludes with recommendations for future research directions in Section \ref{se:conclusion}.

\section{Research Efforts} \label{se:research_efforts}
In this section, we highlight the research results presented at the
ACM Workshop on Wireless Security and Machine Learning (WiseML 2019) that was held in conjunction with the $12^\text{th}$ ACM Conference on Security and Privacy in Wireless and Mobile Networks (ACM WiSec). The goal was to bring together members of the ML, RFML, privacy, security, and wireless communications communities in order for them to share the latest research findings in the emerging and critical area of wireless security and ML. 

The workshop consisted of a keynote, three sessions on ML applications, two sessions on adversarial ML, and two sessions on defense with ML. The keynote \cite{keynote} highlighted the applications of ML in wireless security with examples ranging from conventional attacks formulated with ML to new types of attacks enabled with ML. There were 13 paper presentations and 8 invited talks in three main areas that are discussed in the following three subsections.

\subsection{Adversarial Machine Learning in Wireless Systems} 
Adversarial ML is an emerging field that studies learning in the presence of adversaries \cite{AML1,AML2} and has received major attention in other data domains such as computer vision \cite{Szegedy}. 

In this subsection, we highlight potential uses of adversarial ML in wireless domain.

\begin{itemize}
\item A generative adversarial network (GAN) was used in \cite{1} to generate spoofing wireless signals that cannot be distinguished from real signals. By accounting for channel, waveform, and radio effects, an adversary was shown to train a deep learning based generator that transmits spoofing signals over the air, whereas the defender's goal was to train another deep learning based discriminator to detect spoofing signals. 

\item A GAN was also applied in \cite{2} to generate synthetic RF signals that can be used to confuse spectrum users by jammers, spoofers, and other attackers in wireless radio environments. In particular, GAN models were fit to Long Term Evolution (LTE) and Frequency Modulation (FM) broadcast signals to demonstrate the feasibility of GANs to generate realistic wireless signals.

\item Defense mechanisms were developed in \cite {3} to mitigate  targeted evasion attacks against deep learning-based  RF signal classifiers. Two different datasets were considered, one on modulation recognition and the other one on WiFi 802.11n, Bluetooth and ZigBee classification.

\item Another adversarial example was studied in \cite{4} for targeted attacks against RF signal classifiers. Direct access to the inputs of a deep learning classifier was shown to break down the classifier and the adversarial perturbation power needed to cause source-target misclassification could be used as a proxy for the model's estimation of their similarity.

\item Adaptation of radio parameters such as power control was considered in \cite{5} against deep learning based predictive adversaries in wireless communications. Lyapunov optimization and virtual queues were used to satisfy data transmission reliability in the presence of wireless adversaries while minimizing power consumption.

\item In addition to these papers, \cite{talk1} surveyed research into the threats that adversarial ML poses to deep learning enabled cognitive radios. The primary focus was on evasion attacks by
providing a threat model that characterizes attacks by their prior knowledge, their goals, and the location the attack is launched from. 
\end{itemize}

\subsection{Machine Learning Applications in Wireless Security}
In this subsection, we highlight various applications of  ML techniques in wireless security.
\begin{itemize}
\item Wireless virtualization (WiVi) was discussed in \cite{6} for high speed and quality-of-service (QoS) aware communications while reducing the deployment cost of wireless infrastructure. A Blockchain was used to eliminate over-committing of wireless resources of Wireless Infrastructure Providers such as RF channels, while ML optimally predicts the QoS requirements. 

\item A survey of attacks on ML based wireless systems was presented in \cite{7}. These attacks ranged from model extraction, model inversion and leak of information to evasion and poisoning attacks that aim to manipulate the integrity of test and training data, respectively. 

\item Based on siamese convolutional neural networks (CNNs), a noise-robust signal classification solution was presented in \cite{8}. Trained on compressed spectrogram images, siamese CNNs were shown to effectively distinguish wireless signals by accounting for frequency offsets. 

\item  ML for RF signal processing was discussed in \cite{talk2} in terms of catching the so called ``Third Wave of AI", which refers to the paradigm shift in artificial intelligence from statistical learning and deep neural frameworks, to contextually adaptive, explainable, and generalizable cognitive systems. Recent research programs in Academia, Industry, and Government have been key indicators of the growing interest in applying ML to wireless security where the conceptual framework of the Third Wave is of critical importance. 
 
\item The focus of \cite{talk3} was on how to launch wireless jamming attacks with the assistance of deep learning. The key point was in exploratory attacks that aim to infer the functionality of deep learning classifiers for spectrum sensing followed by subsequent evasion and poisoning attacks to misguide the target classifiers in test and training phases, respectively. 
 
\item The rate and level of performance degradation that occurs when CNNs are subjected to bit-flip errors, which could result from single event upsets that occur in harsh environments, were discussed in \cite{talk4}. The discussion provided a foundation for ongoing research that enhances the overall resilience of neural network architectures and implementations in space under both random and malicious error events, offering significant improvements over current implementations. 
 
\item Initiatives were presented in \cite{talk5} to introduce students to research in RFML applications. Different types of adversarial ML attacks were discussed to highlight threats to cognitive radio systems that increasingly rely on ML to make data-driven automated decisions. 
 
\item Contextual combinatorial bandit learning was discussed in \cite{talk6} for online decision making under uncertainty.  An algorithm was presented to account for issues raised by volatile arms and submodular reward functions and a sublinear regret bound is proven.
 
\item The focus of \cite{talk7} was ML-defined 5G New Radio Networks. Reflecting recent advances in commercial systems, robustness and security were discussed in the application of ML to 5G systems. 
\end{itemize}

\subsection{Wireless Defense with Machine Learning} 
In this subsection, we highlight various cases of defending wireless systems with  ML techniques.
\begin{itemize}
\item Defense strategies were studied in \cite{9} for mobile crowdsensing (MCS) that is vulnerable to illegitimate task injection. ML was shown to eliminate illegitimate tasks with high accuracy and save energy of battery-oriented mobile systems. 

\item A software-defined radio (SDR) implementation was presented in \cite{10} to detect jammers using deep neural networks. Wavelet transform was used as preprocessing to CNN and recurrent neural network (RNN) classifiers that were shown to detect jammers with high accuracy. 

\item Deep learning was applied in \cite{11} to detect adversarial and unintentional communication collisions in a shared spectrum. CNN was used for classification purposes and transfer learning provides the necessary means to scale the training process. 

\item Intrusion detection in IoT systems was studied \cite{12} in the context of anomaly detection. Cumulative sum control chart (CUSUM) was used for timely and accurate anomaly detection followed by in-depth analysis of k-Nearest Neighbors (KNN) distances. 

\item ML was applied in \cite{13} to detect accurately and quickly whether a drone is flying or lying on the ground. Random forest and neural network classifiers were run on features that are based on packet size and inter-arrival time of communications between the drone and its remote controller (RC).

\item A quick and accurate ML-based detection and mitigation solution against IoT-empowered cyber attacks was discussed in \cite{talk8}. The attack space covered volumetric Distributed Denial of Service (DDoS), synchronization-based DDoS, protocol-based DDoS, Byzantine attack, and false data injection attack. 
\end{itemize}

\section{Challenges and Gaps} \label{se:challenges_and_gaps}
\begin{figure}
	\centering
	\includegraphics[width=\columnwidth]{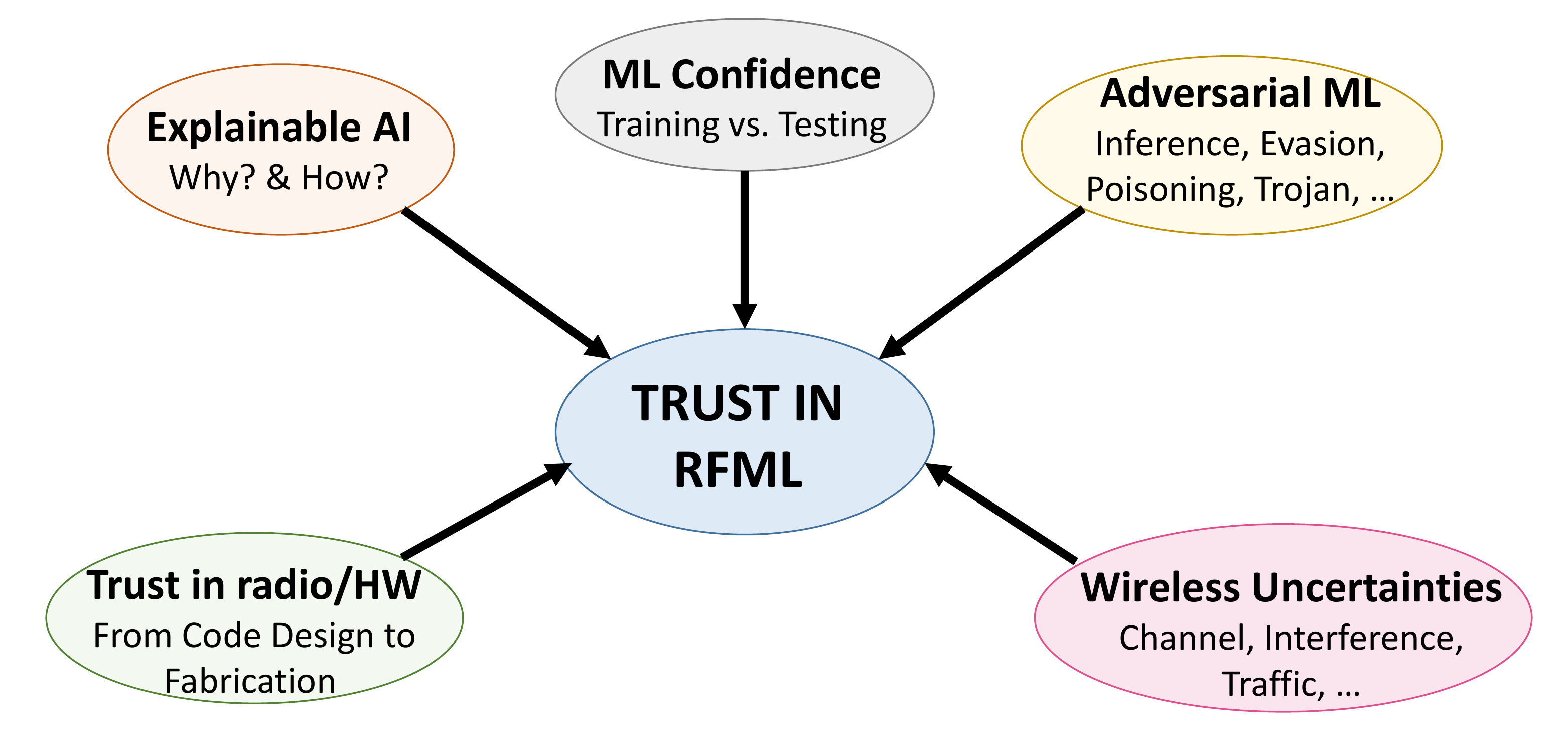}
	\caption{Trust in ML for RFML.}
	\label{fig:fig1}
\end{figure}

One fundamental issue regarding the use of ML in wireless communications and security is whether it is possible to trust ML for these applications. While there are many challenges regarding trust in RFML (with different components shown in Fig.~\ref{fig:fig1}), that are universal (independent of data domain) such as explainability and confidence, some challenges are specific to the wireless domain such as the characteristics of the radio hardware used and the inherent uncertainties associated with the wireless medium.  

\subsection{Development Environment}
One important challenge is how RFML should be deployed. A multi-level development environment will be beneficial while designing RFML solutions that rely on various waveform, channel and radio hardware characteristics. Simulation tests mainly represent virtual hardware and channel effects that may not reflect the real-world scenarios. Simulations need to be followed with emulation tests that include both real traffic and real radios. Emulation tests use actual radios making real transmissions over emulated (virtual) channels \cite{emulation1}. Emulation tests can be repeated by controlling channel effects such as path loss, fading, and delay \cite{emulation2,emulation3}. On the other hand, over-the-air (OTA) testbeds provide the opportunity to test the system with real channels along with real hardware typically in fixed settings.

\begin{figure}[h]
	\centering
	\includegraphics[width=\columnwidth]{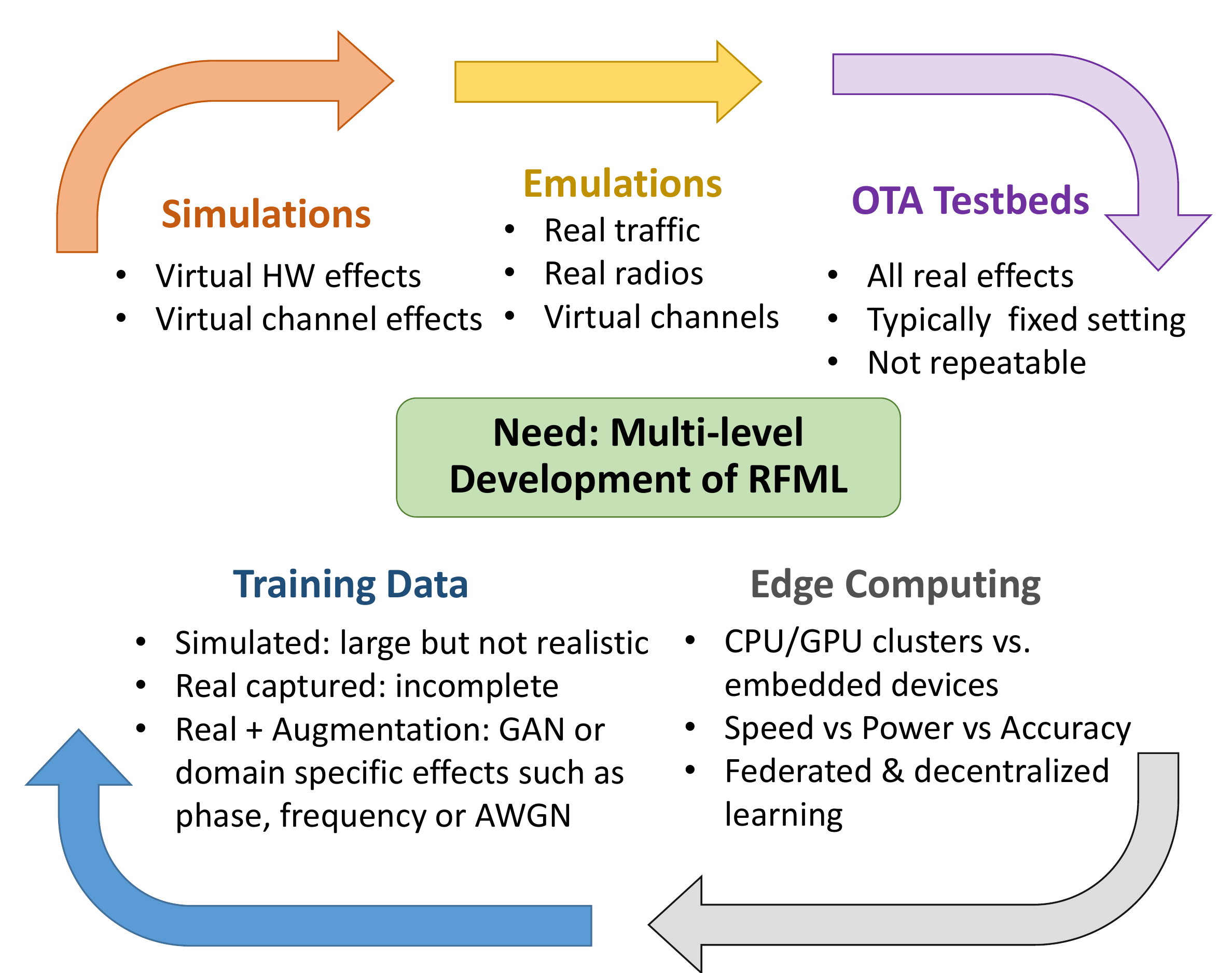}
	\caption{Development environment for RFML.}
	\label{fig:fig2}
\end{figure}

Development environment for RFML should involve necessary training and test datasets, and also reflect potential changes between training and test conditions (e.g., indoor vs. outdoor channel effects) under which data is collected (see Section~\ref{sec:dataset} for more detailed discussion of RFML datasets). In addition, embedded computing should be explored to support edge applications and training data collection for RFML (see Section~\ref{sec:dataset} for more detailed discussion of embedded implementation). Different aspects regarding the need for multi-level development environment for RFML are illustrated in Fig.~\ref{fig:fig2}.

\subsection{Training and Test Datasets} \label{sec:dataset}
ML requires relevant data for both test and training purposes. In wireless security applications, the data needs to reflect the underlying channel, waveform, radio, interference, and adversary  characteristics. While simulated data can be controlled and configured to the needs of the ML problem at hand, it typically lacks fidelity as it is difficult (if not impossible) to model all channel and radio effects reliably. Therefore, real datasets captured over the air are useful resources for ML problems in the wireless domain. 

While there is a trend of releasing real datasets (e.g., see \cite{dataset}), there are not enough datasets available yet to accommodate various scenario needs, e.g., spectrum characteristics in the presence of intelligent jammers. This paradigm raises the danger that research efforts are tailored to the availability of datasets, e.g., there is a plethora of research focused on modulation recognition, where efforts pursuing other signal classification tasks are lagging behind. Therefore, there is a need to produce and publicly release real wireless datasets. In particular, it is of paramount importance to build a wireless security database similar to National Vulnerability Database \cite{NVD} in the cyber domain. 

GAN and other synthetic data generation approaches can be further applied for data augmentation and domain adaptation to satisfy the large data needs of ML algorithms, in particular deep learning. While GAN has been successfully applied to generate synthetic data samples for computer vision \cite{GANGoodfellow} and text analytics  \cite{GANShi} that can be further used to strengthen attacks on target systems \cite{vulnerability}, there is only limited work (e.g., \cite{GAN}) yet to generate synthetic wireless signal samples to support data augmentation and domain adaptation for wireless systems. 

The typical practice is to apply pre-trained ML models to wireless security. Online ML update mechanisms are needed to continue with the optimization of attack and defense space in test time. To that end, reinforcement learning is promising to secure wireless communications in the presence of adversaries such as jammers and eavesdroppers \cite{RL}. ML security in wireless domain differs from its counterpart in other data domains in terms of performance measures. ML can be used to optimize the quality of communications measured by various performance metrics such as throughput and delay subject to power consumption requirements. These metrics are often coupled with each other and need to be jointly optimized \cite{cost}. 

Spectrum is shaped by various effects such as channel, inference, waveform, and traffic \cite{shaping}. Due to the dynamic and random nature of traffic flows and links in a wireless network, it is imperative to maintain queue stability (i.e., queue length remains bounded) \cite{Tassiulas, stability} while optimizing performance via ML. One related performance metric that has gained interest and may benefit from ML is the age of information (AoI) that measures the information freshness (the time elapsed since the last received update) \cite{yates}. 

\subsection{Repeatability, Hyperparameter Optimization, and Explainability}
It is important not only to share the datasets but also the ML code. This will help other researchers quickly extend the proposed solutions to other wireless security problems or compare their own solution to the proposed ones. A common library of algorithmic solutions to ML-based wireless attacks and defenses is ultimately needed. Deep learning has started finding more applications in wireless security as it can capture physical signal characteristics better than conventional ML algorithms such as support vector machines. 

There are four major issues to be solved with deep learning before it can be practically used in wireless applications: 
\begin{enumerate}
\item Deep learning requires large datasets to train. GAN is one way for data augmentation. Additional methods can be borrowed from computer vision and other domains or new ones can be developed to meet this need. 
\item Deep learning has many parameters to optimize such as the number of layers and neurons. Hyperparameter optimization techniques are needed to match deep learning performance to wireless security needs. These techniques should reduce the computational complexity of parameter search. For example, Hyperband \cite{hyperband} iteratively selects and evaluates the performance of hyperparameter configurations with increasing run times. Such techniques reduce the time to select hyperparameters compared to exhaustive search that becomes easily intractable as the neural network structures become deeper. 

\item Deep learning is typically used as a black box. However, wide adoption of wireless security solution requires justification of the performance. For example, preprocessing of spectrum data is crucial to extract features that can be done through statistical, temporal, or spectral methods, and deserves more attention from the research community. In addition, future work should account for the fact that  wireless domain further increases discrepancy between test and training data due to
\begin{itemize}
   \item spectrum dynamics (channel, traffic, interference, and radio effects change over time), 
    \item model changes (data labels (e.g., signal types) may change over time),
    \item noisy data (features, e.g., I/Q samples, are hidden in noise, fading, and interference),
    \item distributed computing (training and inference may be distributed at edge devices), and
    \item adversarial learning (wireless signals are vulnerable to over-the-air attacks based on adversarial ML). 
\end{itemize}

\item In a wireless medium, it is possible that new signal types (such as jammers) may emerge over time. However, it is costly to retrain the signal classifier, whereas accounting  only for new signal types may result in catastrophic forgetting. Continual learning such as using elastic
weight consolidation (EWC) based loss function can be used to strike a balance between old and new tasks to learn. In particular, EWC slows down the learning process on selected ML model parameters to remember
previously learned tasks. In wireless domain, EWC was applied to the problem of wireless signal classification in \cite{dyspan} when the classifier encounters new signal types over time.
\end{enumerate}
\subsection{Embedded Implementation}
The typical practice is offline computation using central processing unit (CPU) or graphics processing unit (GPU)  resources for ML tasks. However, embedded solutions are needed for ML-driven wireless security to support online decision making in size, weight, and power (SWaP) constrained radio platforms. While computational resources such as ARM processors, embedded GPUs and field-programmable gate arrays (FPGA) are available with a wide range of capabilities, research efforts supported by embedded platform implementations are largely missing. Therefore, it is hard to claim whether the current state of the art can match the low latency and low power requirements of wireless security. As examples of embedded implementation of ML solution, ML-based signal authentication was implemented on embedded GPU in \cite{DeepWiFi} and ML-based modulation classification was implemented on Android smartphone in \cite{Smartphone}. 

An alternative implementation is on FPGA. Initial studies showed that FPGA provides major improvements in latency and energy consumption for ML-based RF signal classification compared to embedded GPU \cite{Soltani2019-1, Soltani2019-2}. However, it is known that it is difficult to program FPGAs compared to CPUs and GPUs. Therefore, automated means are needed to convert software codes of ML solutions (especially deep neural networks) to FPGA implementation.  

Table \ref{table:comp} qualitatively compares expected latency, power consumption and programmability of different embedded computing resources. To this end, more efforts should be put to bridge the gap between theory and practice by implementing the proposed solutions on embedded platforms and quantifying latency and energy consumption to run ML solutions.

\begin{table}
\caption{Performance of different of embedded computation platforms.}
\label{table:comp}
\small
\centering
\begin{tabular}{c|c|c|c} \hline \hline \textbf{Measure \textbackslash ~Platform} & \textbf{ARM} & \textbf{Embedded GPU} & \textbf{FPGA}  \\ \hline 
\textbf{Latency} & \emph{High} & \emph{Medium} & \emph{Low} \\ \hline
\textbf{Power Consumption} & \emph{Medium} & \emph{High} & \emph{Low} \\ \hline
\textbf{Programmability} & \emph{High} & \emph{Medium} & \emph{Low} \\ \hline \hline
\end{tabular}
\end{table}

\subsection{Adversarial Machine Learning}
There is a growing interest in applying adversarial ML to the wireless domain. The idea is to attack the training and/or testing processes of ML developed for various application tasks. When applied to wireless domain, such attacks operate with small footprints and are stealthier and more energy efficient compared to conventional attacks that directly target wireless transmissions such as jamming \cite{conventional1, conventional2}, while accounting for uncertainties regarding wireless communications \cite{conventional3, conventional4, conventional5}.

\begin{figure}
	\centering
	\includegraphics[width=\columnwidth]{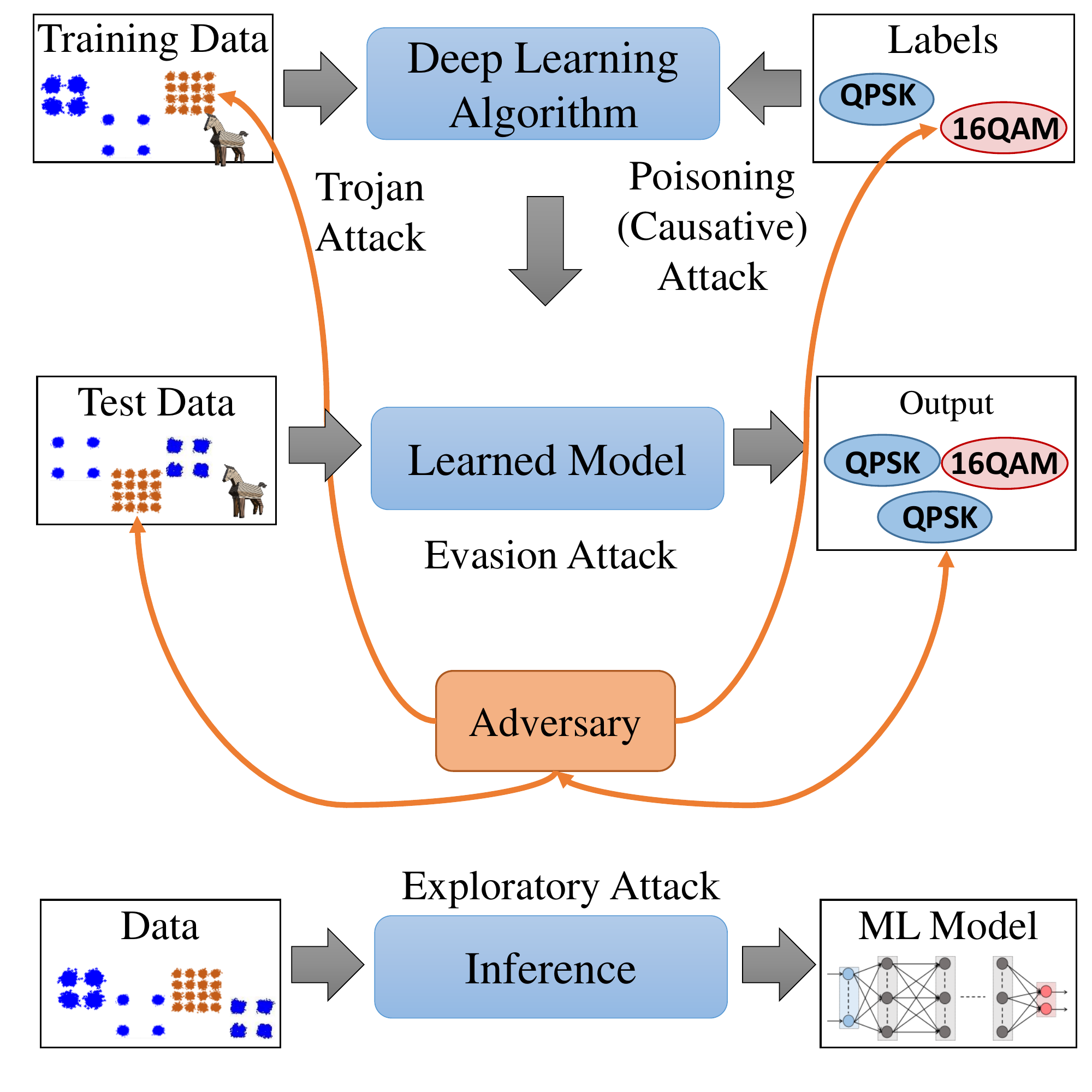}
	\caption{Taxonomy of adversarial ML: evasion, poisoning (causative),  exploratory (inference), Trojan attacks.}
	\label{fig:fig3}
\end{figure}

Different types of adversarial ML are illustrated in Figure~\ref{fig:fig3}. Below, we discuss evasion attacks (adversarial examples),  exploratory (inference), poisoning (causative) attacks, and Trojan attacks in the context of wireless security:  
\begin{itemize}
    \item \emph{Evasion attack}: In an evasion attack, the adversary manipulates the test data to fool a receiver into making errors in classification decisions. Most of the studies in wireless domain have focused on evasion attacks against modulation classifiers \cite{Larsson2018, Headley2019, Deniz2019, Flowers2019, Kim2020}.  Various defense mechanisms are proposed against these attacks \cite{Silvija2019, Silvija2019b}. There are also efforts to apply evasion attacks to other ML-based wireless applications such as spectrum sensing \cite{YiMilcom2018, Yalin2019, AMLTMC2019} and end-to-end communications with an autoencoder \cite{Sadeghi2019} (note that instead of using conventional communication blocks, an autoencoder is trained with two deep neural networks, one as an encoder at the transmitter and the other one as a decoder at the receiver \cite{autoencoder}).
    \item \emph{Exploratory attack}: Exploratory attack aims to infer how an ML algorithm of a victim system works. Typically, this involves building a surrogate model that is functionally equivalent to the victim ML system with the same types of input and outputs \cite{Yi2017}. In wireless domain, exploratory attack was studied in \cite{Yi2018, Tugba2019} to learn transmit patterns of a communications system and build more efficient jamming strategies based on the inferred transmit pattern. Exploratory attack is typically the first step before launching subsequent attacks and benefits from methods like active learning \cite{active} that can help reduce the number of data samples needed to reliably infer the inner workings of an ML algorithm.
    \item \emph{Poisoning attack}: Poisoning attack manipulates the training process of an ML algorithm such that the ML algorithm does not produce outcomes as intended \cite{Yi2017Causative}. Wireless applications include spectrum data falsification to fool classifiers that are built for spectrum sensing \cite{AMLTMC2019, Yalin2019} and cooperative spectrum sensing \cite{Zhuo2019}. These attacks correspond to the ML-generalization of spectrum sensing data falsification (SSDF) attacks that have been extensively studies in the context of cognitive radio security \cite{SSDF}. 
    \item \emph{Trojan attack}: Trojan attack manipulates training data by inserting Trojans (i.e., triggers) to only few training data samples in the training phase and then activates those Trojans in the test phase to fool the ML model into making wrong decisions \cite{TrojanCV}. To selectively fool a wireless signal classifier, a Trojan attack was proposed in \cite{Trojan}, where the adversary modifies phases of few samples and changes corresponding labels in training data, and later transmits signals with the same phase shift that was added as a trigger during training.
\end{itemize}
Adversarial attacks have been also considered in reinforcement learning algorithms with various applications in computer vision \cite{HuangRL}. Studies of adversarial ML in wireless domain are still in their early stages and more work is needed to fully characterize the attack and defense spaces by accounting for unique features of wireless applications such as differences in features and labels observed by adversaries and defenders due to different channel and interference effects.

\section{Conclusion and Recommendations} \label{se:conclusion}
Based on challenges and gaps identified in the previous section, the following conclusions are made:
\begin{itemize}
    \item While ML is critical to secure wireless systems, the attack vectors have not yet been fully understood. More efforts are needed to characterize the vulnerabilities of wireless systems against adversaries that are employing ML techniques. A clearly-defined ML-based attack model is the precondition to formal analysis for ML-related wireless security. 
    \item As adversaries are getting smarter, defense schemes should become more agile and adaptive; ML provides these capabilities and is thus a powerful means to detect and mitigate wireless attacks.
    \item Adversarial ML is important to understand, disrupt or protect the ML process in the presence of wireless adversaries. The wireless medium provides new means for the adversaries to manipulate both test (inference) and training phases of ML. To that end, novel techniques are required to determine the wireless attack surface of adversarial ML. 
    \item The duel between both the attacker and the defender using ML is an evolving, interactive process. Efforts are needed to understand the dynamic evolution of this process with quantification of performance metrics obtained by the attacker and the defender. 
    \item More high-fidelity and diverse datasets that are publicly available are needed to support ML efforts for wireless security. Embedded implementation is crucial to meet latency, power and computational complexity requirements of ML-based attacks and defenses in SWaP-constrained environments.
    
\end{itemize}

\end{document}